%% file: B-L_ConfInf_9.tex
\documentclass[prd,superscriptaddress,twocolumn,nofootinbib,showpacs]{revtex4-1}

\usepackage{amsfonts,amssymb,amsmath}
\usepackage{color}
\usepackage{graphicx}
\usepackage{dcolumn}
\usepackage{bm}

\newcommand{\C}[1]{{\mathcal #1}}

\newcommand{\BF}[1]{{\mathbf #1}}

\newcommand{\beq}{\begin{equation}}
\newcommand{\eeq}{\end{equation}}
\newcommand{\bea}{\begin{eqnarray}}
\newcommand{\eea}{\end{eqnarray}}
\newcommand{\nn}{\nonumber}

\newcommand{\half}{\frac 12}
\newcommand{\third}{\frac 13}
\newcommand{\quarter}{\frac 14}

\newcommand{\Slash}[1]{{\ooalign{\hfil#1\hfil\crcr\raise.167ex\hbox{/}}}}

\begin{document}




\title{Supersymmetric $B-L$ inflation near the conformal coupling}

\author{Masato Arai}
\email{masato.arai(AT)fukushima-nct.ac.jp}
\affiliation{
Institute of Experimental and Applied Physics,
Czech Technical University in Prague,
Horsk\' a 3a/22, 128 00 Prague 2, Czech Republic
}
\affiliation{
Fukushima National College of Technology, Iwaki, Fukushima 970-8034, Japan
}
\author{Shinsuke Kawai}
\email{kawai(AT)skku.edu}
\affiliation{Department of Physics, 
Sungkyunkwan University,
Suwon 440-746, South Korea}
\author{Nobuchika Okada}
\email{okadan(AT)ua.edu}
\affiliation{
Department of Physics and Astronomy, 
University of Alabama, 
Tuscaloosa, AL35487, USA} 

\date{\today}


\begin{abstract}
We investigate a novel scenario of cosmological inflation in 
a gauged $B-L$ extended minimal supersymmetric Standard Model with R-symmetry.
We use a noncanonical K\"{a}hler potential and a superpotential, both preserving the R-symmetry  
to construct a model of slow-roll inflation.
The model is controlled by two real parameters: the nonminimal coupling $\xi$ that originates from
the K\"{a}hler potential, and the breaking scale $v$ of the $U(1)_{B-L}$ symmetry.
We compute the spectrum of the cosmological microwave background radiation and show that
the prediction of the model fits well the recent Planck satellite observation for a wide range of 
the parameter space.
We also find that the typical reheating temperature of the model is low enough to avoid the gravitino
problem but nevertheless allows sufficient production of the baryon asymmetry if we
take into account the effect of resonance enhancement.
The model is free from cosmic strings that impose stringent constraints on generic 
$U(1)_{B-L}$ based scenarios, as in our scenario the $U(1)_{B-L}$ symmetry is broken from the onset.
\end{abstract}



\maketitle


\section{Introduction}\label{sec:Intro}

Recent observations of the cosmological microwave background (CMB) 
\cite{Bennett:2012zja,Hinshaw:2012aka,Ade:2013ktc,Ade:2013uln}
impose stringent restrictions on models of inflation.
For example, the minimally coupled $m^2\phi^2$ and $\lambda\phi^4$ chaotic models that have 
served as simple benchmark models for decades are now in tension.
Inflationary models with nonminimal coupling $\xi\phi^2 R$, where $\phi$ is a scalar field 
(inflaton) and $R$ the scalar curvature, are less constrained, and in fact the predictions of some such 
models have been shown to fit extremely well the current data
\cite{Okada:2010jf}.
It is well known that a nonminimally coupled model (in the Jordan frame) can be Weyl-rescaled to 
a minimally coupled model (the Einstein frame), and hence it is meaningful to discuss the former 
only when the original Lagrangian has significance in its own right, e.g.
when the Lagrangian is that of a particle physics model such as the Standard Model (SM).
The nonminimally coupled SM Higgs inflation 
\cite{Cervantes-Cota1995,Bezrukov:2007ep,DeSimone:2008ei,Bezrukov:2009db,Barvinsky:2009fy}
with the Jordan frame Higgs potential
\beq
V_{\rm J}=\lambda (\phi^2-v^2)^2,
\label{eqn:VJHiggs}
\eeq
provides just such an example.
While the hierarchy problem inherited from the SM and a large dimensionless coupling 
$\xi\sim10^4$ required for the consistency loom large and stay as a matter of debate 
\cite{Burgess:2009ea,Barbon:2009ya,Burgess:2010zq,Bezrukov:2012sa,Salvio:2013rja},
the simplicity and the observational viability are very attractive features.
The hierarchy problem is known to be mitigated in a supersymmetric setup.
Supersymmetric extensions of the Higgs inflation have been proposed
in the next-to-minimal supersymmetric SM \cite{Einhorn:2009bh,Ferrara:2010yw,Ferrara:2010in} 
and in the supersymmetric grand unified theory (GUT) \cite{Arai:2011nq,Pallis:2011gr} 
(see also \cite{Einhorn:2012ih}).
As a closely related model, it is shown in \cite{Arai:2011aa,Arai:2012em} that the Higgs-lepton flat 
direction in the supersymmetric seesaw Lagrangian can realise observationally viable and 
phenomenologically consistent slow roll inflation with small nonminimal coupling $\xi\lesssim{\C O}(1)$.

\begin{table*}[t]
\begin{center}\begin{tabular}{c|cccccccccc}
&~~$Q$~~&~~$u^c$~~&~~$d^c$~~&~~$L$~~&~~$e^c$~~&~~$H_u$~~&~~$H_d$~~&~~
$N^c$~~&~~$S$~~&~~$\Phi_\pm$\\
\hline\\
$SU(3)_c\!\times\! SU(2)_L\!\times\! U(1)_Y$&(${\BF 3},{\BF 2},+\frac 16$)&
($\overline{\BF 3},{\BF 1},-\frac 23$)&($\overline{\BF 3},{\BF 1},+\frac 13$)&
(${\BF 1},{\BF 2},-\frac 12$)&(${\BF 1},{\BF 1},+1$)&(${\BF 1},{\BF 2},+\frac 12$)&
(${\BF 1},{\BF 2},-\frac 12$)&(${\BF 1},{\BF 1},0$)&(${\BF 1},{\BF 1},0$)&(${\BF 1},{\BF 1},0$)\\\\
$U(1)_{B-L}$&$+\third$&$-\third$&$-\third$&$-1$&$+1$&$0$&$0$&$+1$&$0$&$\pm 2$\\\\
$U(1)_{R}$&$+\half$&$+\half$&$+\half$&$0$&$+1$&$+1$&$+1$&$+1$&$+2$&$0$\\\\
\hline
\end{tabular} 
\caption{
The charges of the superfields under the symmetries of the SM gauge group, $U(1)_{B-L}$, 
and $U(1)_{R}$.
\label{table:charges}}
\end{center}
\end{table*}

These popular Higgs inflation models employ positive\footnote{
Our convention is such that the conformal coupling in 4 dimensions corresponds to $\xi=-\frac{1}{6}$.
} nonminimal coupling $\xi$.
Interestingly, it is known that the same Jordan frame potential (\ref{eqn:VJHiggs}) 
with {\em negative} nonminimal coupling $\xi$ also provides a model of inflation
\cite{Linde:2011nh,Kallosh:2013hoa}, in which an observationally viable case correspond to
small field values $|\phi|\lesssim v$.
The potential in the Einstein frame is
\beq
V_{\rm E}\sim\left(\frac{\phi^2-v^2}{1-|\xi|\phi^2}\right)^2,
\label{eqn:VEHiggs}
\eeq
where the reduced Planck mass $M_{\rm P}=(8\pi G)^{-1/2}=2.44\times 10^{18}$ GeV has been
set to unity.
The potential exhibits global minima at $\phi=\pm v$ and singularities at $\phi=\pm1/\sqrt{|\xi|}$.
Hence successful termination of inflation (exit from the slow roll at a global minimum)
requires $v^2|\xi|<1$.
It would be interesting to know in which context of particle physics such a model of inflation may be
implemented, and in particular, what the broken symmetry associated with the potential
(\ref{eqn:VEHiggs}) can be.
Certainly, $\phi$ cannot be the SM Higgs field as the value of $v$ required for
inflation is much larger than the electroweak scale.
A natural guess might be that $\phi$ is a Higgs field responsible for breaking some extra gauge
symmetry.
It would be then important to examine whether the resulting cosmological scenario is 
phenomenologically consistent.

In this paper we point out the possibility that $\phi$ can be the Higgs field associated with 
the $U(1)_{B-L}$ symmetry which is spontaneously broken at an ultra high energy scale
in the early Universe.
The $U(1)_{B-L}$ symmetry is one of the global symmetries of the minimal supersymmetric Standard
Model (MSSM), under which the quark, lepton, and Higgs superfields are charged by
$+\third$, $-1$, and $0$ units (see Table \ref{table:charges}).
We assume that the $U(1)_{B-L}$ gauge symmetry is spontaneously broken at a
super-Planckian scale;
such an assumption is acceptable on phenomenological grounds 
as the breaking scale is experimentally unconstrained except the rather mild LEP bound 
$\gtrsim 3$ TeV \cite{Carena:2004xs,Cacciapaglia:2006pk}.
An important consequence of having the $U(1)_{B-L}$ symmetry is that three generations
of right-handed neutrinos are necessary for anomaly cancellation. 
Thus our model necessarily involves the neutrino sector; this allows us to discuss 
the neutrino masses (via the seesaw mechanism \cite{seesaw}) and the baryon asymmetry of the 
Universe (via thermal \cite{Fukugita:1986hr} or nonthermal \cite{Lazarides:1991wu} leptogenesis) 
within the same model.
Indeed, the $U(1)_{B-L}$-extended SM is one of the leading candidates of the particle physics beyond 
the SM and there are well known inflationary models based on it 
\cite{Copeland:1994vg,Dvali:1994ms,Lazarides:1995vr,Lazarides:1996dv,Jeannerot:1997av,Jeannerot:2000sv,Jeannerot:2001qu,Senoguz:2005bc,Rehman:2009nq,Buchmuller:2013lra}
(see also \cite{Okada:2011en,Okada:2013vxa}).
The novelty of our scenario, in comparison to the existing ones, is simplicity of the construction
and robustness of the prediction.
Also, our model is free from overproduction of cosmic strings that generally afflicts the 
$U(1)_{B-L}$-based inflationary models; in our scenario the $U(1)_{B-L}$ symmetry is already 
broken at the onset of inflation and there is no danger of producing topological defects during and 
after inflation.
We shall consider supergravity-embedding
as the nonminimal coupling of the inflaton naturally arises in such a framework.

The rest of the paper is organised as follows.
In the next section we construct the model from the supergravity setup, and in Sec. \ref{sec:dynamics} 
we discuss the inflationary dynamics.
We compare the prediction of this model with the results of the Planck satellite observation in 
Sec. \ref{sec:spectrum}, and the post-inflationary physics is discussed in Sec. \ref{sec:RH}. 
We conclude the paper with comments in Sec. \ref{sec:concl}. %

\section{The model}\label{sec:model}

The starting point of our model is the superpotential
\beq
W_{\rm eff}=\kappa S(\Phi_+\Phi_- -v^2),
\label{eqn:Weff}
\eeq
where the superfields $S$, $\Phi_+$, $\Phi_-$ are singlets in the SM
gauge group $SU(3)_{\rm c}\times SU(2)_{L}\times U(1)_Y$, and carry 
$0$, $+2$, $-2$ units of $U(1)_{B-L}$ charges.
We choose $\kappa>0$, $v>0$ by field redefinition.
The local $U(1)_{B-L}$ symmetry is broken by the vacuum expectation value
$v$ of the $\Phi_\pm$ fields.
The model may be considered as a part of a supersymmetric SM whose superpotential is 
(for example)
\bea
W&=&\mu H_u H_d+y_u^{ij} u_i^c Q_j H_u +y_d^{ij} d_i^c Q_j H_d +y_e^{ij} e_i^c L_j H_d\nn\\
&+&\kappa S(\Phi_-\Phi_+-v^2)+y_D^{ij}N_i^c L_j H_u+ \lambda^{ij} \Phi_- N_i^cN_j^c.
\label{eqn:W}
\eea
The first line represents the MSSM and the last two terms are responsible for the seesaw mechanism and leptogenesis.
Here, $Q$, $u^c$, $d^c$, $L$, $e^c$, $H_u$, $H_d$ are the MSSM superfields,
$N^c$ the right-handed neutrino superfields, $y$'s are the Yukawa couplings and $\mu$ is the MSSM
$\mu$ parameter.
The family indices are $i, j={1, 2, 3}$.
There are three right-handed neutrinos necessary for anomaly cancellation.
The Majorana Yukawa coupling $\lambda^{ij}$ controls the seesaw scale; we shall discuss it
in a later section.
With $+2$, $0$, $0$ units of R-charges assigned to the $S$, $\Phi_+$, $\Phi_-$ fields, the 
superpotential is also invariant under the $U(1)_R$ symmetry\footnote{
These $R$-charges are what is called $R'=R-\half L$ in \cite{Lazarides:1998iq}.
In our model both $R$ and $R'$ are conserved.}.
In Table \ref{table:charges} we list the SM gauge group, $U(1)_{B-L}$ and $U(1)_{R}$ charges
assigned to the superfields appearing in the superpotential (\ref{eqn:W}).
Note that (\ref{eqn:Weff}) is the most general renormalisable superpotential for $S$ and $\Phi_\pm$ 
that is compatible with these symmetries. 
For supergravity embedding in the superconformal framework, we shall use a slightly noncanonical 
K\"{a}hler potential $K=-3\Phi$, where
\bea
\Phi&=&1-\third\left(|\Phi_+|^2+|\Phi_-|^2+|S|^2\right)\nn\\
&&\qquad+\frac{\gamma}{2}\left(\Phi_+\Phi_-+\Phi_+^*\Phi_-^*\right)
+\frac{\zeta}{3}|S|^4.
\label{eqn:Phi}
\eea
This preserves the $U(1)_R$ symmetry and contains two real parameters $\gamma$ and $\zeta$.

The scalar potential is found by the standard supergravity computation \cite{sugra}. 
We take the D-flatness direction $|\Phi_+|=|\Phi_-|\equiv\half\varphi$ and define
\beq
\Phi_+=\half\varphi e^{i\theta_+},\quad
\Phi_-=\half\varphi e^{i\theta_-},
\eeq
with real scalar fields $\varphi$, $\theta_\pm$.
The F-term scalar potential is then\footnote{
The super-Planckian inflaton values imply that 
higher dimensional operators are not negligible. 
To avoid deformation of the potential due to such
operators, some degree of fine-tuning is unavoidable.}
\bea
V_{\rm F}
&=&
\half\kappa^2|S|^2\varphi^2
+\frac{\kappa^2}{1-4\zeta |S|^2}\left(\frac{\varphi^4}{16}-\half\varphi^2 v^2\cos\theta+v^4\right)\nn\\
&&\hspace{-8mm}-\kappa^2|S|^2\frac{
\left|\left(\frac 34 \gamma-\frac{\zeta |S|^2 e^{i\theta}}{1-4\zeta |S|^2}\right)\frac{\varphi^2}{2}
-2\left(1-\frac{\zeta|S|^2}{1-4\zeta |S|^2}\right)v^2\right|^2}{3-\frac{3\gamma}{8}\varphi^2\cos\theta
+\frac{9\gamma^2}{32}\varphi^2
+\frac{\zeta |S|^4}{1-4\zeta |S|^2}},
\label{eqn:VF}
\eea
where $\theta\equiv\theta_++\theta_-$.
Note that $V_{\rm F}$ is invariant under the phase shift $S\rightarrow e^{i\alpha} S$, reflecting
the unbroken $U(1)_{R}$ symmetry.

\begin{figure}
\includegraphics[width=80mm]{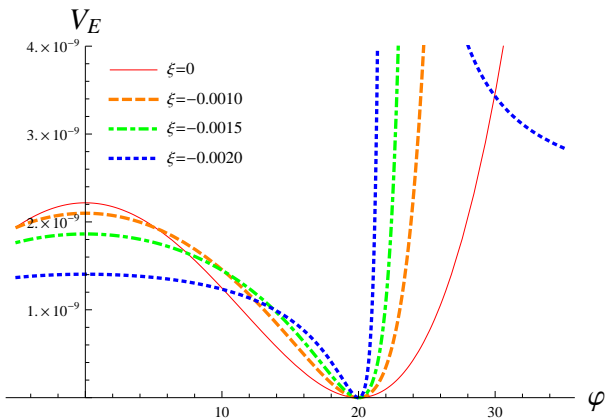}
\caption{\label{fig:potential}
The scalar potential in the Einstein frame for $v=10$ and $\xi=0, -0.0010, -0.0015, -0.0020$.
The value of $\kappa$ is determined by the amplitude of the primordial density perturbation (see text).
The e-folding number is chosen to be $N_e=60$.
}
\end{figure}

\begin{table*}[t]
\begin{center}\begin{tabular}{c|ccccccccccccc}
${\xi}$&$\kappa^2$ &~& $\varphi_*$ &~& $\varphi_k$ &~& $n_s$ &~& $dn_s/d\ln k$ &~& $r$ &~&
$\delta$\\
\hline
0 & ~$2.22\times 10^{-13}$ && 18.6 && 7.05 && 0.964 &&$-4.54\times 10^{-4}$ && 0.0519 &&1\\
$-0.0010$~ & ~$2.10\times 10^{-13}$ && 19.0 && 9.44 && 0.968 &&$-4.65\times 10^{-4}$ && 0.0466 &&0.6\\
$-0.0015$~ & ~$1.86\times 10^{-13}$ && 19.2 && 11.1 && 0.970 &&$-4.63\times 10^{-4}$ && 0.0408 &&0.4\\
$-0.0020$~ & ~$1.40\times 10^{-13}$ && 19.4 && 13.5 && 0.972 &&$-4.56\times 10^{-4}$ && 0.0313 &&0.2\\
\end{tabular} 
\caption{The parameter $\kappa$, the inflaton field value at the end of the slow roll $\varphi_*$ and at the horizon exit of the comoving CMB scale $\varphi_k$, the spectral index $n_s$ and its running $dn_s/d\ln k$,
the tensor-to-scalar ratio $r$ and the parameter $\delta$ defined in (\ref{eqn:delta}) computed in the example of 
$v=10$ and $\xi=0, -0.0010, -0.0015, -0.0020$.
We have chosen the e-folding number $N_e=60$.
\label{table:v=10}}
\end{center}
\end{table*}

Let us comment on the F-term hybrid inflation (FHI) models
\cite{Copeland:1994vg,Dvali:1994ms,Lazarides:1995vr,Lazarides:1996dv,Jeannerot:1997av,Jeannerot:2000sv,Jeannerot:2001qu,Senoguz:2005bc}
which share some similarity with ours.
The FHI models are based on the same superpotential (\ref{eqn:Weff}) representing the spontaneously
broken local $U(1)_{B-L}$ symmetry, but usually the canonical K\"{a}hler potential
is assumed\footnote{There are FHI models with a noncanonical K\"{a}hler potential, including
\cite{BasteroGil:2006cm,Garbrecht:2006az,urRehman:2006hu,Pallis:2009pq,Rehman:2010wm,Armillis:2012bs}.}.
In these models the major role is played by the $S$ field whereas the role played by $\varphi$ 
(called a waterfall field) is minor.
At the tree level the scalar spectral index $n_s$ of the FHI models is typically enhanced (blue).
The slightly red $n_s$ compatible with the current observations can be obtained by including radiative 
and supergravity correction terms (see \cite{Pallis:2013dxa,Orani:2013rra} for up-to-date accounts).
Here in our model we take a different trajectory from the FHI models.
It can be shown that for large enough $\zeta$, the field $S$ is stabilised at $S=0$ and its dynamics 
can be neglected (cf. \cite{Ferrara:2010yw,Ferrara:2010in}).
Then the potential (\ref{eqn:VF}) and $\Phi$ of (\ref{eqn:Phi}) simplify to
\bea
V_{\rm F}&=&\frac{\kappa^2}{16}\left(\varphi^4-8\varphi^2 v^2\cos\theta+16 v^4\right),\\
\Phi&=& 1+\left(\frac{\gamma}{4}\cos\theta-\frac{1}{6}\right)\varphi^2.
\eea
Examining the F-term scalar potential in the Einstein frame $V_{\rm E}=\Phi^{-2}V_{\rm F}$, 
it can be checked that the phase is stable at $\theta=0$ 
(we will be concerned with the parameter region $\gamma>0$; see below). 
Thus one may further ignore the dynamics of $\theta$. 
The system then reduces to a single field model for which the scalar-gravity part of the
Lagrangian is written as
\beq
{\C L}_{\rm J}=
\sqrt{-g_{\rm J}}\Big[
\half\Phi R_{\rm J}-\half g_{\rm J}^{\mu\nu}\partial_\mu\varphi\partial_\nu\varphi
-V_{\rm J}\Big],
\eeq
where the subscript J stands for the Jordan frame and
\bea
\Phi&=&1+\xi\varphi^2,\\
V_{\rm J}&=&V_{\rm F}=\kappa^2\left(\frac{\varphi^2}{4}-v^2\right)^2,\\
\xi&=&\frac{\gamma}{4}-\frac 16.
\eea
The Lagrangian in the Einstein frame is related to the one in the Jordan frame by the 
Weyl transformation 
$g^{\rm E}_{\mu\nu}=\Phi g^{\rm J}_{\mu\nu}$ and is written as
\beq
{\C L}_{\rm E}
=\sqrt{-g_{\rm E}}\Big[
\half R_{\rm E}-\half g_{\rm E}^{\mu\nu}\partial_\mu\hat\varphi\partial_\nu\hat\varphi-V_{\rm E}\Big],
\label{eqn:LagE}
\eeq
where $R_{\rm E}$ is the scalar curvature in the Einstein frame, 
$g_{\rm E}^{\mu\nu}=(g^{\rm E}_{\mu\nu})^{-1}$, 
$V_{\rm E}=\Phi^{-2}V_{\rm J}$, 
and $\hat\varphi$ is the canonically normalised scalar fields in the Einstein frame which is
related to $\varphi$ via
\beq
d\hat\varphi=\frac{\sqrt{1+\xi \varphi^2+6\xi^2 \varphi^2}}{1+\xi \varphi^2} d\varphi.
\eeq
The slow roll parameters defined in the Einstein frame are
\bea
&&\epsilon_V=\half\left(\frac{1}{V_{\rm E}}\frac{d V_{\rm E}}{d\hat\varphi}\right)^2,
\qquad 
\eta_V=\frac{1}{V_{\rm E}}\frac{d^2 V_{\rm E}}{d\hat\varphi^2},\nn\\
&&
\xi_V^2=\frac{1}{V_{\rm E}^2}\frac{d V_{\rm E}}{d\hat\varphi}\frac{d^3 V_{\rm E}}{d\hat\varphi^3}.
\eea
Using the original field $\varphi$ these are expressed as
\bea
\epsilon_V&=&\half\left(\frac{V'_{\rm E}}{V_{\rm E}\hat\varphi'}\right)^2,
\qquad
\eta_V=\frac{V''_{\rm E}}{V_{\rm E}(\hat\varphi')^2}
-\frac{V'_{\rm E}\hat\varphi''}{V_{\rm E}(\hat\varphi')^3},
\\
\xi_V^2&=&\frac{V'_{\rm E}}{V_{\rm E}^2}\left(
3\frac{V'_{\rm E}(\hat\varphi'')^2}{(\hat\varphi')^6}
-\frac{V'_{\rm E}\hat\varphi'''}{(\hat\varphi')^5}
-3\frac{V''_{\rm E}\hat\varphi''}{(\hat\varphi')^5}
+\frac{V'''_{\rm E}}{(\hat\varphi')^4}\right),\nn
\eea
where the prime means ${}'\equiv d/d\varphi$.

\section{Inflationary dynamics}\label{sec:dynamics}

In the last section we obtained the single field inflationary model from the gauged
$B-L$ extended MSSM.
The inflaton potential in the Einstein frame is
\beq
V_{\rm E}=\frac{\kappa^2}{16}\left(\frac{\varphi^2-4v^2}{1+\xi \varphi^2}\right)^2,
\label{eqn:VEeff}
\eeq
which is essentially the one (\ref{eqn:VEHiggs}) discussed in the introduction.
The potential (\ref{eqn:VEeff}) has supersymmetric vacua at $\varphi=\pm 2v$ and
singularities at $\varphi=\pm1/\sqrt{|\xi|}$.
Without losing generality we shall focus on positive $v$ and positive $\varphi$.
In this paper we will be interested in the inflationary scenario with negative $\xi$ 
\cite{Linde:2011nh,Kallosh:2013hoa}.
The initial value of the inflaton is between $0<\varphi<2v$, namely, between the local maximum of 
the potential $V_{\rm E}(0)=\kappa^2 v^4$ and the supersymmetric vacuum at $\varphi=2v$.
This is analogous to the {\em new inflation} model, or even newer, hilltop type models 
\cite{Boubekeur:2005zm}.
For successful termination of the slow roll the supersymmetric vacuum $\varphi=2 v$ should not be
hidden behind a singularity; this requires $2v<1/\sqrt{|\xi|}$. 
The physics beyond the singularity is of no interest to us, as it is the antigravity regime
where the Newton constant becomes negative \cite{Linde:1979kf,1981SvAL....7...36S}.

In our model there are three tunable parameters $\kappa$, $v$ and $\xi$.
These are constrained by the amplitude of the density perturbation of the comoving CMB scale.
For definiteness we use the maximum likelihood value
$A_s(k_0)=2.215\times 10^{-9}$ from the Planck satellite observation \cite{Ade:2013ktc} 
with the pivot scale at $k_0=0.05 \mbox{ Mpc}^{-1}$.
With this $A_s(k)$ the power spectrum of the curvature perturbation
${\C P}_R=V_{\rm E}/24\pi^2\epsilon_V$ at the horizon exit of the comoving scale is normalised as $A_s(k)=\frac{k^3}{2\pi^2}{\C P}_R(k)$.
The end of the slow roll is characterised by the condition that one of the slow roll parameters
that are small during inflation becomes ${\C O}(1)$.
We obtain the inflaton value at the end of the slow roll 
$\varphi_*$ by solving $\max(\epsilon, |\eta|)= 1$, and then find the inflaton value 
$\varphi_k$ at the horizon exit of the comoving CMB scale $k$ by solving 
$N_{e}=\int_{\varphi_*}^{\varphi_k}d\varphi V_{\rm E}({d\hat\varphi}/{d\varphi})/({d V_{\rm E}}/{d\hat\varphi})$
for an e-folding number $N_e$. 
In this way the value of $\kappa$ is fixed once $N_e$, $v$ and $\xi$ are given.

The inflaton potential (\ref{eqn:VEeff}) includes various cases in its limits \cite{Linde:2011nh}.
When $\xi\rightarrow 0$, $v\rightarrow 0$ it approaches to the minimally coupled 
$\lambda\phi^4$ model,
while the limit $\xi\rightarrow 0$, $v\rightarrow\infty$ gives the prediction obtained in the 
minimally coupled $m^2\phi^2$ model.
It is also known that the $4|\xi|v^2\rightarrow 1$ limit yields the same inflationary prediction as the 
nonminimally coupled Higgs inflation with large positive $\xi$.
As an indication of how close to this limit our model is, we introduce a parameter
\beq
\delta\equiv 1-4|\xi|v^2.
\label{eqn:delta}
\eeq
This is actually $\Phi$ of (\ref{eqn:Phi}) appearing in the K\"{a}hler potential, evaluated at the
supersymmetric vacuum at $\varphi=2v$ and $S=0$.
In the limit $\xi\rightarrow 0$ the potential becomes the double-well type;
the prediction of this inflationary model is compatible with the combined Planck$+$WP$+$BAO results 
\cite{Ade:2013uln} when $2v\gtrsim 13$ in our parametrisation at 95\% confidence level (CL).
As $|\xi|$ is increased 
the singularity at $\varphi=1/\sqrt{|\xi|}$ approaches the potential minimum at $\varphi=2v$.
Fig. \ref{fig:potential} shows the shape of the potential when $v=10$ and 
$\xi=0$, $-0.0010$, $-0.0015$ and $-0.0020$.
The scalar spectral index $n_s\equiv 1+d\ln A_s(k)/d\ln k\simeq 1-6\epsilon_V+2\eta_V$,
the running of the spectral index ${dn_s}/{d\ln k}\simeq -24\epsilon_V^2+16\epsilon_V\eta_V-2\xi_V^2$,
and the tensor-to-scalar ratio $r\equiv {\C P}_{\rm gw}/{\C P}_{\rm R}\simeq16\epsilon_V$ 
are evaluated by computing the slow roll parameters at the horizon exit of the comoving CMB scale.
We list these results for $v=10$ and the e-folding number $N_e=60$ in Table \ref{table:v=10},
along with the values of $\kappa$, $\varphi_*$ and $\varphi_k$ found by the procedure explained 
above.
The nonminimal coupling is varied as $\xi=0$, $-0.0010$, $-0.0015$, 
$-0.0020$.
The tendency of these CMB parameters may be understood from the behaviour of the potential
in Fig. \ref{fig:potential}.
As $|\xi|$ is increased, the minimum of the potential becomes a steep valley, while the small
$\varphi$ region becomes a plateau; consequently, the spectrum of the inflationary model approaches 
to that of the nonminimally coupled Higgs inflation model.

\section{Comparison with Planck}\label{sec:spectrum}

\begin{figure*}
\includegraphics[width=89mm]{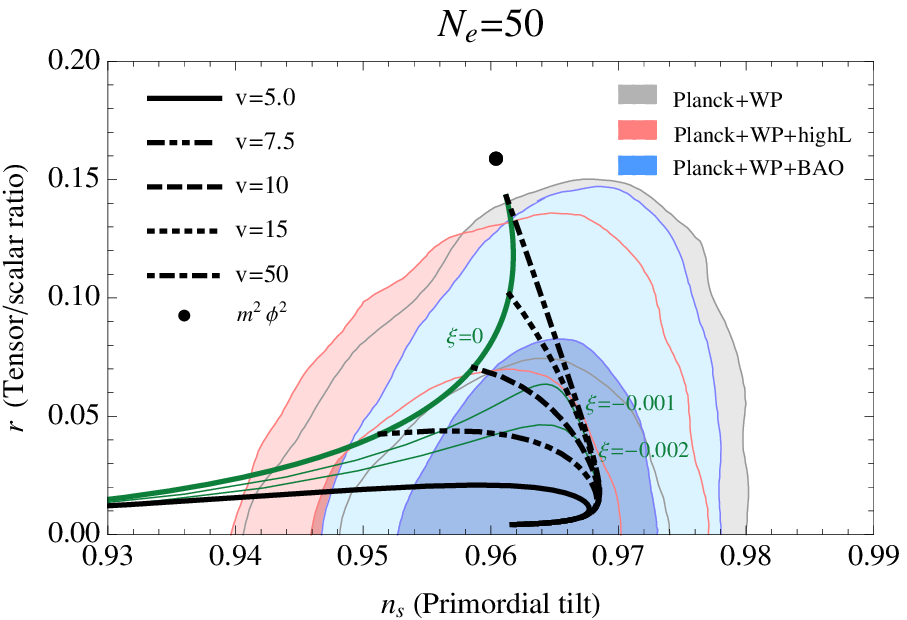}
\includegraphics[width=89mm]{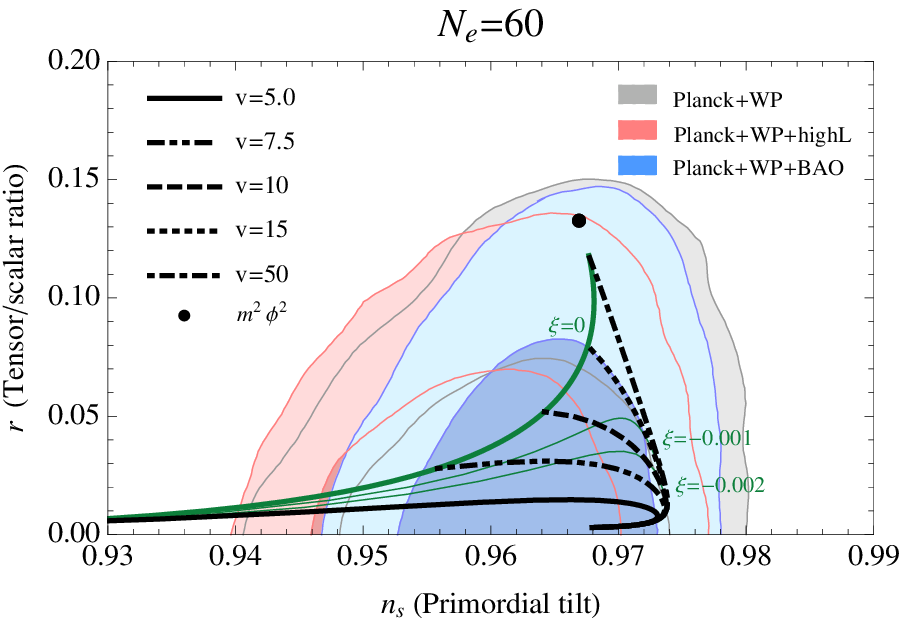}
\caption{\label{fig:rns_Planck}
The spectral index $n_s$ and the tensor-to-scalar ratio $r$ of our model, when
the symmetry breaking parameter $v$ and the value of the nonminimal coupling $\xi$ are varied 
as $v=5.0, 7.5, 10, 15, 50$ and $-1/4v^2<\xi\leq0$.
The left (right) panel shows the results for the e-folding number $N_e=50$ ($N_e=60$).
The 68\% and 95\% CL contours from the Planck satellite observation
\cite{Ade:2013ktc}
(Planck$+$WP: grey, Planck$+$WP$+$highL: red, Planck$+$WP$+$BAO: blue, from the background
to the foreground) 
are shown for comparison.
The black dot is the prediction of the minimally coupled $m^2\phi^2$ chaotic inflation model, i.e.
$n_s=1-2/(N_e+\half)$ and $r=4(1-n_s)$, to which our model approaches in the limit $\xi=0$, 
$v\rightarrow\infty$.
The thick green curve is $\xi=0$ and the thin green curves are $\xi=-0.001$ and $\xi=-0.002$
from above.
}
\end{figure*}

In this section we show the prediction of our model for the scalar spectral index $n_s$, the running of 
the scalar spectral index $d n_s/ d\ln k$, and the tensor-to-scalar ratio $r$ for varying $v$.
The nonminimal coupling is varied as $-1/4v^2<\xi\leq0$. 
The procedure and the normalisation are as described in the previous section.

Fig.\ref{fig:rns_Planck} shows the plots of $n_s$-$r$, the left panel showing the results for 
$N_e=50$ and the right panel for $N_e=60$.
The 68\% and 95 \% contours from the Planck satellite observation \cite{Ade:2013ktc}
(Planck$+$WP: grey, Planck$+$WP$+$highL: red, Planck$+$WP$+$BAO: blue)
are superimposed on the background for comparison.
In the minimally coupled case ($\xi=0$) small $v$ is strongly disfavoured 
($v\gtrsim 6.5 M_{\rm P}$ for Planck$+$WP$+$BAO 95\% CL \cite{Ade:2013uln}).
With small negative $\xi$, in contrast, we see that smaller $v$ (see $v=5$ for example)
is not only compatible but in excellent fit with the current CMB data. 
This feature is favourable for the model as the large super-Planckian excursion of the inflaton is 
often considered problematic.

\begin{figure*}
\includegraphics[width=89mm]{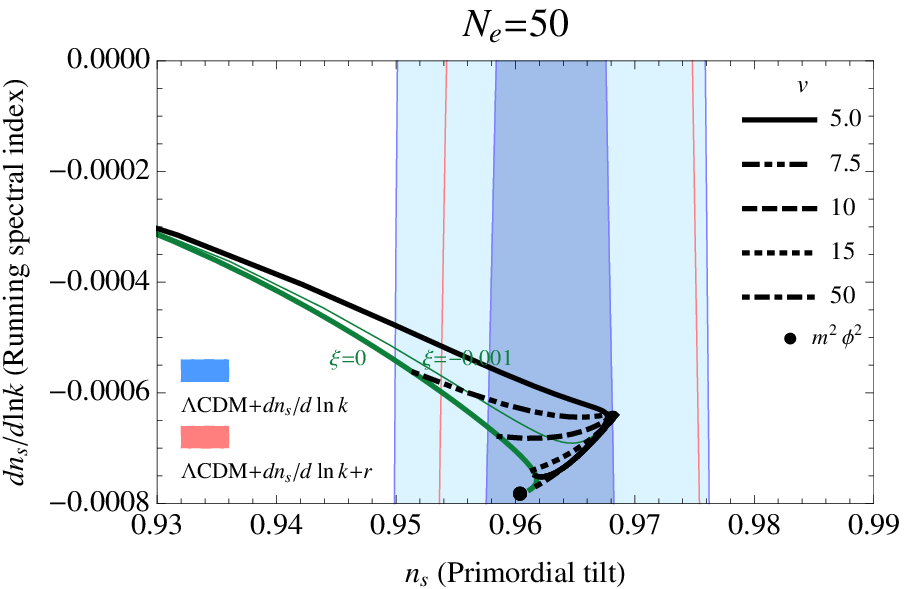}
\includegraphics[width=89mm]{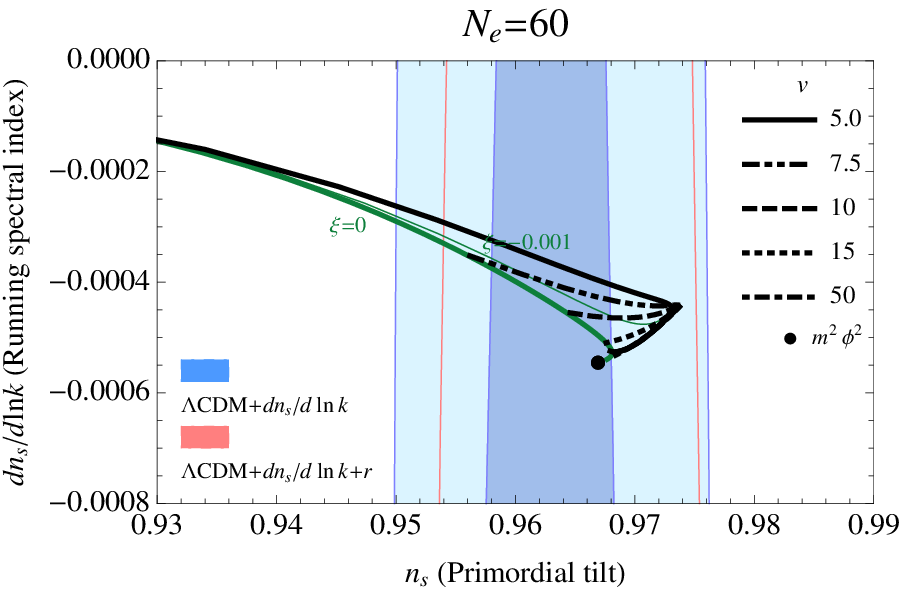}
\caption{\label{fig:nsrun_Planck}
The prediction of our model for the running of the scalar spectral index $dn_s/d\ln k$ against the 
scalar spectral index $n_s$.
The left panel shows the results for the e-folding number $N_e=50$ and the right panel is for 
$N_e=60$.
The parameters are the same as in Fig. \ref{fig:rns_Planck}. 
The 68\% and 95\% CL contours of the Planck$+$WP$+$BAO \cite{Ade:2013uln} 
are indicated in the background.
The 68\% and 95\% CL contours of $\Lambda$CDM$+dn_s/d\ln k$ are shown by blue and light blue, and the 95\% CL contour of $\Lambda$CDM$+dn_s/d\ln k+r$ is shown by the thin red curves 
(the 68\% CL contour is outside the figure).
The black dot is the prediction of the minimally coupled $m^2\phi^2$ chaotic inflation model, 
namely 
$n_s=1-2/(N_e+\half)$ and $dn_s/d\ln k=-2/(N_e+\half)^2$, to which our model approaches in the 
$\xi=0$, $v\rightarrow\infty$ limit.
The thick (thin) green curve is the prediction for $\xi=0$ ($\xi=-0.001$).
}
\end{figure*}

The running of the scalar spectral index $dn_s/d\ln k$ is shown against the scalar spectral index $n_s$ 
in Fig. \ref{fig:nsrun_Planck} (the left panel: $N_e=50$, the right panel: $N_e=60$).
The 68\% and 95\% CL contours of the Planck$+$WP$+$BAO \cite{Ade:2013uln} are also shown for 
comparison.
In the figure the 68\% and 95\% CL contours of $\Lambda$CDM$+dn_s/d\ln k$ are shown by dark and light blue, and the 95\% CL contour of $\Lambda$CDM$+dn_s/d\ln k+r$ is shown by the light red curve.
The 68\% CL contour of $\Lambda$CDM$+dn_s/d\ln k+r$ is outside the figure.
While the running of the scalar spectral index is potentially an important observable beyond 
$n_s$ and $r$, the data at present is not significant enough to restrict the model parameters;
the contours run nearly vertical in the figure, indicating that the constraints are mainly due to $n_s$.

%
Going back to Fig.\ref{fig:rns_Planck}, we see that the prediction of our model for $n_s$ and $r$
makes stark contrast to the nonminimally coupled $\lambda\phi^4$ model 
(see e.g. \cite{Okada:2010jf,Arai:2011aa,Arai:2012em}) in which the prediction moves
vertically in the $n_s$-$r$ plane as the nonminimal coupling $\xi$ is varied.
In Fig.\ref{fig:rns_Planck} the parameter space of our model covers almost the whole area inside
the 68\% CL contour; it would be interesting to see how future observations
\cite{Hazumi:2008zz,Bock:2009xw,Bouchet:2011ck,Kermish:2012eh}, in particular precision 
measurements of $n_s$, will constrain these parameters.

\section{Reheating and leptogenesis}\label{sec:RH}

In this section we discuss 
viability of the post-inflationary physics.
Our scenario is based on the well-motivated particle physics model of the gauged $B-L$ extended
MSSM with R-symmetry, which has been studied in detail.
A peculiar feature of our model is that the breaking scale $v\approx {\C O}(10)\times M_{\rm P}$ is 
large, compared to the
GUT (see e.g. \cite{,Senoguz:2005bc,Buchmuller:2012wn}) or the electroweak scale
(e.g. \cite{Iso:2009ss,Iso:2009nw,Iso:2010mv,Burell:2011wh}) $B-L$ breaking scenarios.

\begin{figure}
\includegraphics[width=87mm]{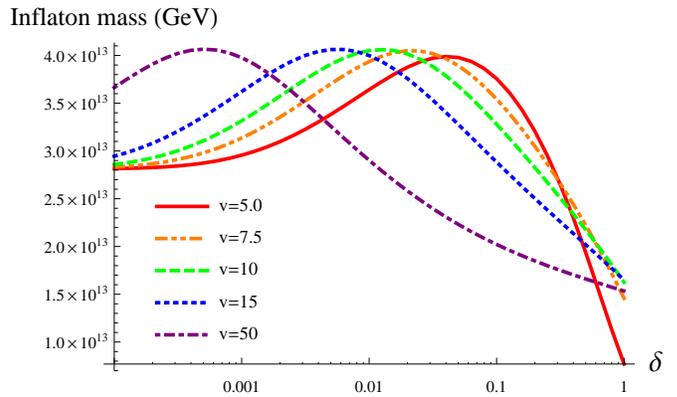}
\caption{\label{fig:InflatonMass}
The inflaton mass $m_{\delta\hat\varphi}$, plotted against $\delta$ (\ref{eqn:delta})
for $v=5.0$, $7.5$, $10$, $15$, $50$. 
}
\end{figure}

Assuming perturbative decay of the inflaton, the upper bound of the reheating temperature is 
estimated as
\beq
T_{\rm RH}\lesssim
\left(\frac{90}{g_*\pi^2}\right)^\quarter\sqrt{M_{\rm P}\Gamma_{\delta\hat\varphi}},
\label{eqn:TRH}
\eeq
where $g_*\approx 200$ is the degrees of freedom at reheating and $\Gamma_{\delta\hat\varphi}$
is the decay rate of the inflaton in the Einstein frame $\delta\hat\varphi$ oscillating at the potential 
minimum. 
The mass of the inflaton is 
\beq
m_{\delta\hat\varphi}^2=\frac{\partial^2 V_{\rm E}}{\partial\hat{\varphi}^2}\Big|_{\varphi=2 v}
=\frac{2\kappa^2 v^2}{1+4\xi v^2(1+6\xi)},
\eeq
which is found to be $m_{\delta\hat\varphi}=10^{13}$-$10^{14}$ GeV for our model 
parameters (Fig.\ref{fig:InflatonMass}).
We are interested in the decay of the inflaton to the SM particles.
The inflaton is a component of $\Phi_\pm$ and there are two pertinent channels of decay: \\

(i) $\delta\hat\varphi\rightarrow NN$, $\widetilde{N}\widetilde{N}$
(the right-handed (s)neutrinos).\\

(ii) $\delta\hat\varphi\rightarrow Z' Z'$ (the $U(1)_{B-L}$ gauge bosons).\\\\
Let us consider the decay channel (i) first.
The decay in this case is through the coupling $\lambda\equiv\lambda^{ij}$ 
in the last term of (\ref{eqn:W}).
After inflation the field $\Phi_-$ settles down at $\langle\Phi_-\rangle =v$, giving the seesaw scale 
$M_N=\lambda\langle\Phi_-\rangle=\lambda v$.
Using the neutrino mass
$m_\nu^2\approx\Delta m_{32}^2=2.43\times 10^{-3}\mbox{ eV}^2$  \cite{Fogli:2012ua} and the
Higgs expectation value $\langle H_u\rangle\approx 174$ GeV in the 
seesaw relation 
$m_\nu M_N=y_D^2\langle H_u\rangle^2$, we see from $y_D\lesssim{\C O}(1)$ that
the seesaw scale is bounded from above: 
$M_N\lesssim 10^{13}$ GeV.
Since $v\approx {\C O}(10)\times M_{\rm P}$ in our model, the Majorana Yukawa coupling needs to 
be small, 
$\lambda\lesssim 10^{-6}$.
The decay rate of the inflaton is
\beq
\Gamma(\delta\hat\varphi\rightarrow NN, \widetilde{N}\widetilde{N})
\approx\frac{\lambda^2}{16\pi}m_{\delta\hat\varphi}.
\eeq
Using the seesaw relation and $m_{\delta\hat\varphi}\approx 10^{13}$ GeV,
we find the reheating temperature from (\ref{eqn:TRH}),
\beq
T_{\rm RH}\approx \frac{M_N}{10^{12} \mbox{ GeV}}\times 10^{7}\mbox{ GeV}.
\label{eqn:Trh1}
\eeq
The second channel (ii) also contributes when the $Z'$ mass is smaller than (half of) the inflaton
mass $m_{\delta\hat\varphi}$.
From the longitudinal mode dominant decay width
\beq
\Gamma(\delta\hat\varphi\rightarrow Z'Z')
\approx\frac{1}{32\pi}\frac{m_{\delta\hat\varphi}^3}{v^2},
\eeq
the reheating temperature is estimated as
\beq
T_{\rm RH}\approx 10^{6} \mbox{ GeV}.
\label{eqn:Trh2}
\eeq
The case (i) may be regarded as the dominant channel.
The condition that the Big Bang nucleosynthesis is not spoiled by the thermally produced gravitinos 
yields an upper bound of the reheating temperature $T_{\rm RH}\lesssim 10^6$-$10^7$ GeV
\cite{Kawasaki:2004yh,Kawasaki:2004qu}.
By (\ref{eqn:Trh1}), this gravitino constraint mildly restricts the seesaw scale $M_N\lesssim 10^{12}$ GeV.

In the out-of-equilibrium decay of the right-handed (s)neutrinos, lepton asymmetry can be generated
and is later converted to the baryon asymmetry of the Universe via the sphaleron transitions,
the so-called leptogenesis scenario \cite{Fukugita:1986hr,Lazarides:1991wu}.
In the sphaleron transition the yield (the ratio of the number density to the entropy density) of the 
baryons is related to that of the leptons as 
\beq
Y_B\approx-\frac{8}{23}Y_L.
\eeq
In {\em thermal leptogenesis} scenario in which the right-handed (s)neutrinos are thermally produced,
the reheating temperature needs to be higher than the mass scale of the right-handed (s)neutrinos:
$T_{\rm RH}\gtrsim M_N$.
The generated baryon asymmetry is estimated as
\cite{Buchmuller:2002rq,Buchmuller:2004nz,Buchmuller:2005eh}
\beq
Y_B\sim Y_L\sim a\frac{\varepsilon_i}{g_*},
\label{eqn:TLG}
\eeq
where $\varepsilon_i$ is the CP asymmetry parameter associated with the $i$'th generation of the 
right-handed (s)neutrino $N_i$ ($\widetilde{N}_i$), and $a\lesssim 1$ is the efficiency factor which 
depends on details of the Boltzmann equations.
The baryon asymmetry of the Universe is observed to be \cite{Hinshaw:2012aka,Ade:2013zuv}
\beq
Y_B=(8.55\pm 0.217)\times 10^{-11} \quad (\mbox{95\% CL}),
\label{eqn:BAU}
\eeq
and thus the CP asymmetry parameter needs to be $\varepsilon_i\approx 10^{-7}$.
This condition is known to be satisfied when the right-handed (s)neutrino mass is large enough, 
$M_{N_i}\gtrsim 10^{10}$ GeV \cite{Buchmuller:2002rq,Buchmuller:2004nz}.
In our scenario, however, this requirement cannot be fulfilled as the reheating temperature (\ref{eqn:Trh1}), (\ref{eqn:Trh2}) is not high enough to produce such heavy right-handed (s)neutrinos.
Nevertheless, it is known that even if the right-handed (s)neutrino masses are not large,
large enough $\varepsilon_i$ (and thus $Y_B$) can be obtained when at least two of the right-handed 
(s)neutrino masses are nearly degenerate and resonant enhancement takes place 
\cite{Flanz:1996fb,Pilaftsis:1997jf}.
Thus, we may think of the following two possible cases of thermal leptogenesis in our scenario:
(a) when the decay channel (i) is dominant, the reheating temperature is given by (\ref{eqn:Trh1})
and thus one of the right-handed (s)neutrinos masses needs to be large.
For successful resonant thermal leptogenesis the remaining two masses need to be very close to each other
and less than the reheating temperature, for example, 
$M_{N_1}\approx M_{N_2}\ll T_{\rm RH} \ll M _{N_3}$;
(b) when the reheating temperature is determined by the decay channel (ii) and is given by 
(\ref{eqn:Trh2}),
all $N_i$'s can be light: $M_{N_1}$, $M_{N_2}$, $M_{N_3}\ll T_{\rm RH}$.
In both (a) and (b), the right-handed (s)neutrinos that produce lepton number through the decay 
are much lighter than $10^{10}$ GeV and hence the resonance enhancement needs to take place.
For detailed analysis of resonant leptogenesis in the context of the minimal B-L model, see, for 
example \cite{Iso:2010mv,Okada:2012fs}.

In {\em nonthermal leptogenesis} scenario, the seesaw scale $M_N$ is larger than the reheating 
temperature $T_{\rm RH}$ and the right-handed (s)neutrinos are predominantly produced by the 
decay process (i).
The baryon asymmetry generated by the decay of the (s)neutrinos
may be estimated as \cite{Buchmuller:2005eh}
\beq
Y_B\sim Y_L\sim \frac{T_{\rm RH}}{m_{\delta\hat\varphi}}
\sum_i {\rm Br}_i \varepsilon_i,
\label{eqn:NTLG}
\eeq
where
\beq
{\rm Br}_i=\frac{\Gamma(\delta\hat\varphi\rightarrow N_iN_i,
\widetilde{N}_i\widetilde{N}_i)}
{\Gamma_{\delta\hat\varphi}}
\eeq
is the branching ratio and $\Gamma_{\delta\hat\varphi}$ is the total decay width of the inflaton.
Using $T_{\rm RH}\sim 10^{7}$ GeV and $m_{\delta\hat\varphi}\approx 10^{13}$ GeV
in (\ref{eqn:NTLG}) we find ${\rm Br}_i\varepsilon_i\approx 10^{-4}$.
This large CP asymmetry parameter can be again obtained by resonant leptogenesis.
To conclude, for the production of the baryon asymmetry through thermal or nonthermal leptogenesis
the CP asymmetry parameter needs to be $\varepsilon_i\sim10^{-7}$-$10^{-4}$.
These values are somewhat larger than the conventional decay scenarios of the right-handed
(s)neutrinos, but can be accounted for by the resonance enhancement 
\cite{Flanz:1996fb,Pilaftsis:1997jf}.
For this, at least two of the right-handed neutrino masses need to be nearly degenerate.
The baryon asymmetry may, alternatively, be generated by some other mechanism such as the 
Affleck-Dine mechanism \cite{Affleck:1984fy}.

\section{Discussions}\label{sec:concl}

In this paper we have constructed a novel scenario of cosmological inflation based on the 
gauged $B-L$ extended MSSM.
The model is well-motivated by the neutrino physics: the spontaneous breaking of the 
$U(1)_{B-L}$ symmetry gives rise to the right-handed neutrino mass term, which in turn give the small nonzero
left-handed neutrino masses through the type I seesaw mechanism.
It also includes the mechanism of baryogenesis through leptogenesis.
Due to supersymmetry our model is stable against radiative corrections and includes
supersymmetric particles that may be considered a good dark matter candidate.
Our model has various advantages over the popular FHI models.
For example the observationally supported slightly red spectrum of the primordial density perturbation
can be naturally accounted for.
There is no need to invoke radiative corrections, and thus there is no necessity of
fine-tuning associated with it.

A notable feature of our scenario is that it is free from unwanted relic particles.
As discussed in Sec.\ref{sec:RH} the reheating temperature is low enough so that the gravitino 
problem can be avoided.
In addition, our model is free from cosmic strings that generally impose stringent constraints
on the FHI models (see e.g. \cite{Lazarides:1995vr}). 
This is due to our assumption that the 
$U(1)_{B-L}$ symmetry is already broken at the onset of inflation and cosmic strings are inflated away.

We conclude the paper by commenting on issues that are of potential importance but
were not discussed above. 
The original model (\ref{eqn:Weff}) includes multiple fields 
and we simplified the model by focusing on the $\varphi$ direction.
This is justified by assuming the quartic term in the K\"{a}hler 
potential that controls the tachyonic instability (see \cite{Ferrara:2010yw,Ferrara:2010in}).
Lifting this assumption certainly complicates the scenario, but  
leads to rich observable consequences. 
The resulting model is sensitive to the initial condition of the inflaton field, as in the case of the
FHI scenario. 
The extra degrees of freedom give rise to the isocurvature mode and possibly non-Gaussianity
of the density perturbation (see e.g. \cite{Mazumdar:2010sa} for a review).
If such a signature is to be detected in the future, it will be an indication that the single field 
approximation that we used is clearly inappropriate.
Our second comment concerns the interpretation of the $\delta\rightarrow 0$ limit.
The $\delta$ defined in (\ref{eqn:delta}) is the factor $\Phi$ of the K\"{a}hler potential (\ref{eqn:Phi})
at the potential minimum.
There is no reason for it to be extremely large or small, and thus the ``Higgs-inflation limit" 
$\delta\rightarrow 0$ is a limit of fine-tuning. 
In this sense, this $\delta\rightarrow 0$ limit is not much better than the $\xi\gg 1$ Higgs inflation.
In Fig.\ref{fig:rns_Planck}, $\delta=1$ ($\xi=0$) is indicated by the thick green curve and smaller
$\delta$ (larger $|\xi|$) corresponds to smaller $r$. 
See also Table \ref{table:v=10}.
The CMB polarisation experiments 
\cite{Hazumi:2008zz,Bock:2009xw,Bouchet:2011ck,Kermish:2012eh} are expected to uncover the
physics of primordial gravitational waves, with accuracy corresponding to $r\sim 0.01$.
These experiments will tell us whether one actually needs to consider the fine-tuned small $\delta$ limit. 
Finally we comment on possible extensions of the model.
The essential elements of our model are the superpotential (\ref{eqn:Weff}) 
and the K\"{a}hler potential (\ref{eqn:Phi}), and hence, it is easy to construct a similar inflaton 
potential if a supersymmetric model is equipped with the same structure.
It is, however, not straightforward to construct a consistent scenario of inflation since keeping the R-symmetry in a realistic GUT is known to be extremely difficult \cite{Fallbacher:2011xg}.

\subsection*{Note added}
After completion of this paper detection of the CMB B-mode polarisation 
was announced by the BICEP2 experiment \cite{Ade:2014xna}, with the tensor/scalar ratio
$r=0.20^{+0.07}_{-0.05}$.
This means in our model that small values of $|\xi|$, and hence $\delta\sim 1$ {\em without} 
fine-tuning (see \eqref{eqn:delta}), are favoured.

\subsection*{Acknowledgments}
This work was supported in part by the Grants-in-Aid for Scientific Research from the
Ministry of Education, Culture, Sports, Science, and Technology of Japan (No.25400280),
the Research Program MSM6840770029 and the project of International Cooperation ATLAS-CERN 
LG13009 of the Ministry of Education, Youth and Sports of the Czech Republic (M.A.), 
by the National Research Foundation of Korea Grant-in-Aid for Scientific Research 
No. 2013028565 
(S.K.)
and by the DOE Grant No. DE-FG02-10ER41714 (N.O.).
A part of the numerical computation was carried out using computing facilities at the
Yukawa Institute, Kyoto University.

\bigskip


\input{B-L_ConfInf_9.bbl}

\end{document}

%% file: B-L_ConfInf_9.bbl
%